\documentclass[twocolumn,journal]{IEEEtran}

\usepackage{amsfonts}
\usepackage{amsmath}
\usepackage{amsthm}
\usepackage{amssymb}
\usepackage{graphicx}
\usepackage[T1]{fontenc}
\usepackage{supertabular}
\usepackage{longtable}
\usepackage[usenames,dvipsnames]{color}
\usepackage{bbm}
\usepackage{fancyhdr}
\usepackage{breqn}
\usepackage{url}
\usepackage{capt-of}
\usepackage{tikz}
\usetikzlibrary{matrix}
\usepackage{endnotes}
\usepackage{soul}
\usepackage{marginnote}
\newcommand{\mathsym}[1]{}
\newcommand{\unicode}[1]{}

\usepackage{enumerate}
\usepackage{colortbl}

\usepackage[framemethod=TikZ]{mdframed}
\usepackage[framemethod=TikZ]{mdframed}
\usepackage{framed}

\mdfsetup{%
	nobreak=true,
	outerlinewidth=1,skipabove=20pt,backgroundcolor=yellow!30, outerlinecolor=black,innertopmargin=0pt,splittopskip=\topskip,skipbelow=\baselineskip, skipabove=\baselineskip,ntheorem,roundcorner=5pt}

\mdtheorem[nobreak=true,outerlinewidth=1,
backgroundcolor=yellow!50, outerlinecolor=black,innertopmargin=0pt,splittopskip=\topskip,skipbelow=\baselineskip, skipabove=\baselineskip,ntheorem,roundcorner=5pt,font=\itshape]{result}{Result}

\mdtheorem[nobreak=true,outerlinewidth=1,
backgroundcolor=yellow!30, outerlinecolor=black,innertopmargin=0pt,splittopskip=\topskip,skipbelow=\baselineskip, skipabove=\baselineskip,ntheorem,roundcorner=5pt,font=\itshape]{theorem}{Theorem}

\mdtheorem[nobreak=true,outerlinewidth=1,
backgroundcolor=gray!10, outerlinecolor=black,innertopmargin=0pt,splittopskip=\topskip,skipbelow=\baselineskip, skipabove=\baselineskip,ntheorem,roundcorner=5pt,font=\itshape]{remark}{Remark}

\mdtheorem[nobreak=true,outerlinewidth=1,
backgroundcolor=gray!10, outerlinecolor=gray!10,innertopmargin=0pt,splittopskip=\topskip,skipbelow=\baselineskip, skipabove=\baselineskip,ntheorem,roundcorner=5pt,font=\itshape]{definition}{Definition}

\mdtheorem[nobreak=true,outerlinewidth=1,
backgroundcolor=pink!30, outerlinecolor=black,innertopmargin=0pt,splittopskip=\topskip,skipbelow=\baselineskip, skipabove=\baselineskip,ntheorem,roundcorner=5pt,font=\itshape]{quaestio}{Quaestio}

\mdtheorem[nobreak=true,outerlinewidth=1,
backgroundcolor=yellow!50, outerlinecolor=black,innertopmargin=5pt,splittopskip=\topskip,skipbelow=\baselineskip, skipabove=\baselineskip,ntheorem,roundcorner=5pt,font=\itshape]{background}{Background}

\mdtheorem[nobreak=true,outerlinewidth=1,
backgroundcolor=gray!10, outerlinecolor=black,innertopmargin=5pt,splittopskip=\topskip,skipbelow=\baselineskip, skipabove=\baselineskip,ntheorem,roundcorner=5pt,font=\itshape]{nothing}{}

\mdtheorem[nobreak=true,outerlinewidth=1,
backgroundcolor=pink!50, outerlinecolor=black,innertopmargin=5pt,splittopskip=\topskip,skipbelow=\baselineskip, skipabove=\baselineskip,ntheorem,roundcorner=5pt,font=\itshape]{point}{Point}
\mdtheorem[nobreak=true,outerlinewidth=1,
backgroundcolor=pink!50, outerlinecolor=black,innertopmargin=5pt,splittopskip=\topskip,skipbelow=\baselineskip, skipabove=\baselineskip,ntheorem,roundcorner=5pt,font=\itshape]{lemma}{Lemma}

\mdtheorem[nobreak=true,outerlinewidth=1,
backgroundcolor=pink!50, outerlinecolor=black,innertopmargin=5pt,splittopskip=\topskip,skipbelow=\baselineskip, skipabove=\baselineskip,ntheorem,roundcorner=5pt,font=\itshape]{commentary}{Comment}
\title{\color{Brown} 
On Single Point Forecasts for Fat-Tailed Variables
}

\author{
    \IEEEauthorblockN{Nassim Nicholas Taleb\IEEEauthorrefmark{1}\IEEEauthorrefmark{2}, Yaneer Bar-Yam\IEEEauthorrefmark{3}, and Pasquale Cirillo\IEEEauthorrefmark{4}\IEEEauthorrefmark{5} 
    } \\
    
        \IEEEauthorblockA{\IEEEauthorrefmark{1}Universa Investments}, 
        \IEEEauthorblockA{\IEEEauthorrefmark{2}Tandon School of Engineering, New York University}\\
    \IEEEauthorblockA{\IEEEauthorrefmark{3}New England Complex Systems Institute}\\
    \IEEEauthorblockA{\IEEEauthorrefmark{4}Institute for the Future, University of Nicosia}, 
     \IEEEauthorblockA{\IEEEauthorrefmark{5}S-T-A-T-S GmbH, Switzerland}\\
 { \color{Maroon} Accepted, \textit{International Journal of Forecasting}}
    \thanks{July 23, 2020; nnt1@nyu.edu}
    
}

\begin{document}
\maketitle
{\color{Brown}
\begin{abstract}
We discuss common errors and fallacies when using naive "evidence based" empiricism and point forecasts for fat-tailed variables, as well as the insufficiency of using naive first-order scientific methods for tail risk management. 

We use the COVID-19 pandemic as the background for the discussion and as an example of a phenomenon characterized by a multiplicative nature, and what mitigating policies must result from the statistical properties and associated risks. In doing so, we also respond to the points raised by Ioannidis et al.(2020)
	
\end{abstract}}
  \thanks{   July 27, 2020. We thank Pierre Pinson and Spyros Makridakis for helping organize this discussion, and John Ioannidis for his gracious engagement.  }

\begin{mdframed}
\smallskip
\section*{\color{Brown}Main Statements}
\begin{enumerate}[(i)--]
	\item Forecasting single variables in fat-tailed domains is in violation of both common sense and probability theory. 
	\item Pandemics are extremely fat-tailed events, with potentially destructive tail risk. Any model ignoring this is necessarily flawed.
	\item Science is not about making single points predictions but about understanding properties (which can \textit{sometimes} be tested by single point estimates and predictions).
	\item Sound risk management is concerned with extremes, tails and their full properties, and not with averages, the bulk of a distribution or naive estimates. 
	\item Naive fortune-cookie evidentiary methods fail to work under both risk management and fat tails, because the absence of raw evidence can play a large role in the properties.
	\item There are feedback mechanisms between forecast and reaction that affects the validity of some predictions. 
	\item Individuals risks fail to translate into systemic risks under multiplicative processes.
	\item One should never treat the "costs" of mitigation without taking into account the costs of the disease, and in some cases naive cost-benefit analyses fail (for sure when statistical averages are nonconvergent or invalid for tail risk purposes).
\end{enumerate}
\smallskip
\end{mdframed}

\begin{mdframed}
\smallskip
\begin{enumerate}
	\item[(ix)] Historically, in the aftermath of the Great Plague, economies were less fragile to pandemics, equipped to factor-in effective mechanisms of containment (quarantines) in their operating costs. It is more cogent to blame overoptimization than reaction to disease.
\end{enumerate}
\smallskip

\end{mdframed}

The article is organized at three levels. First, we make general comments around the nine points in the Main Statements, explaining how single point forecasts is an unscientific simplification incompatible with processes with richer properties. Next we go deeper into the technical arguments. Finally we address specific points in Ioannidis et al. \cite{Ioannidisblog} and answer their arguments concerning our piece.
\begin{figure}
\includegraphics[width=\linewidth]{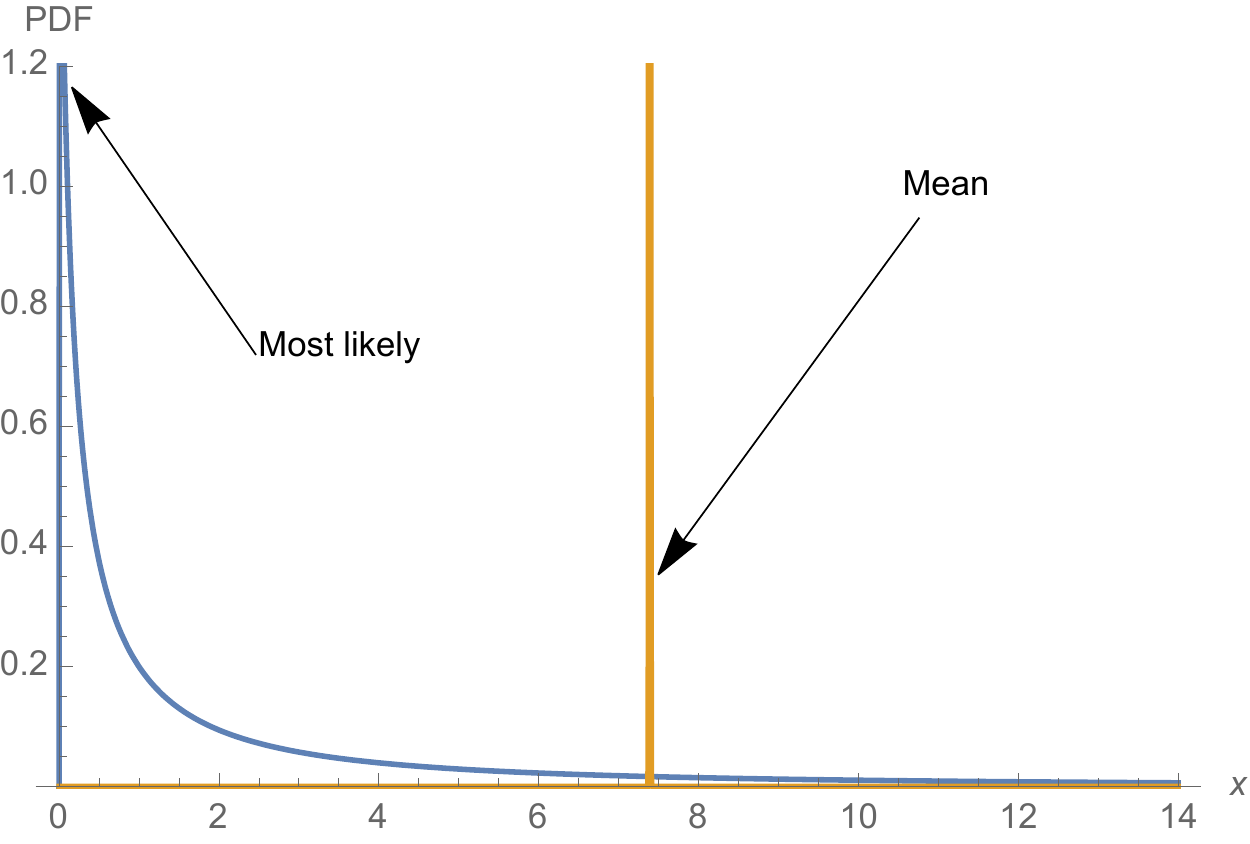}
 \caption{A high variance Lognormal distributions. 85\% of observations fall below the mean. Half the observations fall below 13\% of the mean. The lognormal has milder tails than the Pareto which has been shown to represent pandemics.}\label{lognormal}
\end{figure}

\begin{figure}
\includegraphics[width=\linewidth]{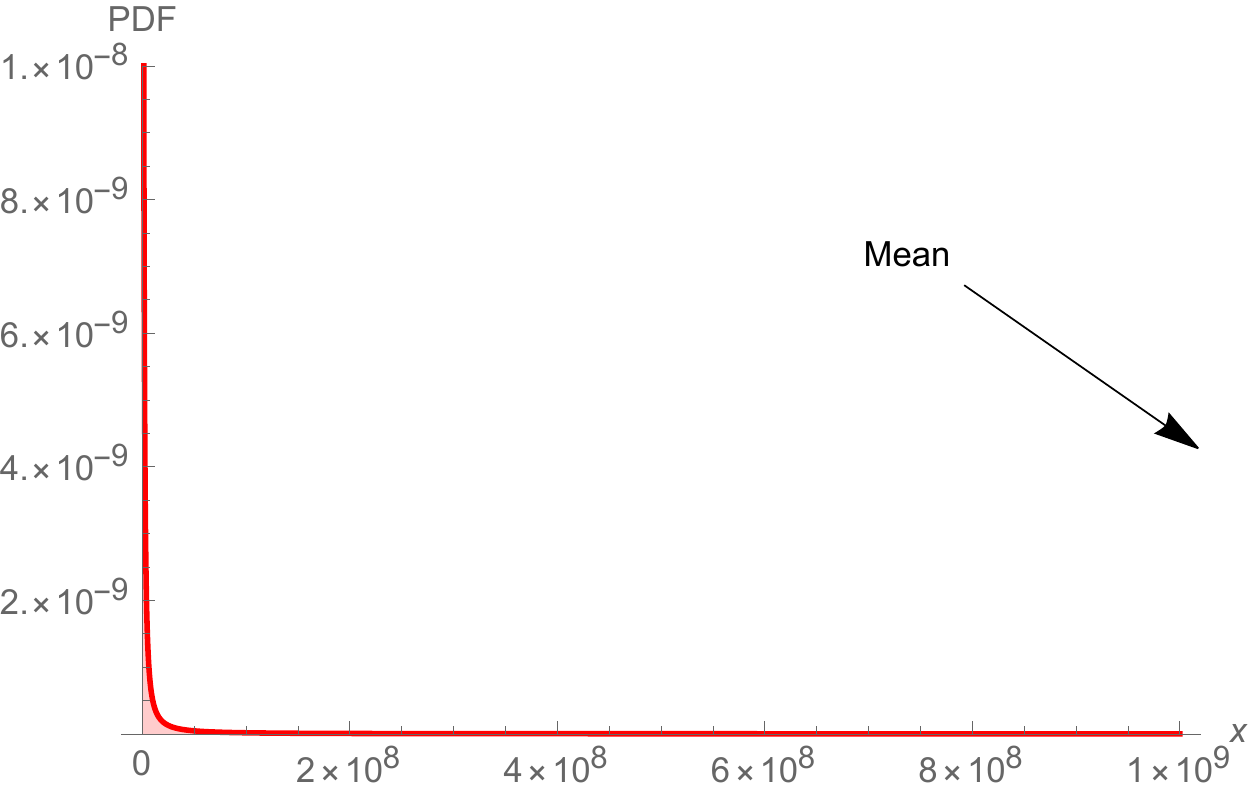}
 \caption{A Pareto distribution with a tail similar to that of the pandemics. Makes no sense to forecast a single point. The "mean" is so far away you \textit{almost} never observe it. You need to forecast things other than the mean. And most of the density is where there is noise.} \label{pareto}
\end{figure}
\section*{\color{Brown}Commentary}

\subsubsection*{\color{Brown}Both forecasters and their critics are wrong}
At the onset of the COVID-19 pandemic, many research groups and agencies produced single point "forecasts" for the pandemic---most relied upon trivial logistic regressions, or upon the compartmental SIR model, sometimes supplemented with cellular automata, or with agent-based models assuming various social rules and behaviors. Apparently, the prevailing idea is that \textit{producing a single numerical estimate} is how science is done, and how science-informed decision-making ought to be done: bean counters producing precise numbers. And always within a narrowly considered set of options identified by the researchers.

Well, no. That is not how "science is done", at least in this domain, and that is not how informed decision-making should develop.

Furthermore, subsequently and ironically, many criticized the plethora of predictions produced, because these did not play out (no surprise there). This is also wrong, because both forecasters (who missed) and their critics (complaining) were wrong. Indeed, forecasters would have been wrong anyway, even if they had got their predictions right. In fact, as we will clarify throughout this article, 1) in some domains (i.e. under fat tails) naive forecasts are poor descriptors of a system (hence highly unscientific), even when they might appear reasonable; 2) for some functions (risk management related), or some classes of exposures (systemic ones), these forecasts are extremely misplaced.

\subsubsection*{\color{Brown}Statistical attributes of pandemics} 
Using tools from extreme value theory (EVT), Cirillo and Taleb \cite{CiTaNat} have recently shown that pandemic deaths are \textit{patently} fat-tailed\footnote{A non-negative continuous random variable $X$ has a fat-tailed distribution, if its survival function $S(x)=P(X\geq x)$ is regularly varying, formally $S(x)=L(x) x^{-\alpha}$, where $L(x)$ is a slowly varying function, for which $\lim_{x \to \infty}\frac{L(tx)}{L(x)}=1$ for $t>0$ \cite{deHaan,Embrechts,Falk}. The parameter $\alpha$ is known as the tail parameter, and it governs the fatness of the tail (the smaller $\alpha$ the fatter the tail) and the existence of moments ($E[X^p]<\infty$ if and only if $\alpha>p$).}--a fact some people like Benoit Mandelbrot (or one of the authors, in \textit{The Black Swan \cite{Incerto}}) had already guessed, but never formally investigated. Even more, the estimated tail parameter $\alpha$ is smaller than 1, suggesting an apparently infinite risk \cite{CiTaNat}, in line with destructive events like wars \cite{CiTa,CiTaQF}, and the so-called "dismal" theorem \cite{Weiz}. Pandemics do therefore represent a source of existential risk. The implication is that much of what takes place in the bulk of the distribution is just noise, according to "the tail wags the dog" effect \cite{CiTaNat,TaStat}. And one should never forecast, pontificate, or theorise from noise! Under fat tails, all relevant and vital information lies in fact in the tails themselves (hence in the extremes), which can show remarkably stable properties.

\begin{remark}[Observed events vs Observed Properties]
 Random variables with unstable (and uninformative) sample moments may still have extremely stable and informative tail properties, centrally useful for robust inference and risk taking. Furthermore, these reveal evidence.
 
This is the central problem with the misunderstanding of \textbf{The Black Swan} \cite{Incerto}: some events may have stable and well-known properties, yet they do not lend themselves to prediction.
\end{remark}

\subsubsection*{\color{Brown} Fortune-cookie evidentiary methods} In the early stages of the COVID-19 pandemic, scholars like Ioannidis \cite{Ioannidis1} suggested that one should wait for "more evidence" before acting with respect to that pandemic, claiming that "we are making decisions without reliable data". 

Firstly, there seems to be some probabilistic confusion, leading towards the so-called delay fallacy \cite{Hansson}: "if we wait we will know more about X, hence no decision about X should be made now." 

In front of potentially fat-tailed random variables, more evidence is not necessarily needed. Extra (usually imprecise) observations, especially when coming from the bulk of the distribution, will not guarantee extra knowledge. Extremes are rare by definition, and when they manifest themselves it is often too late to intervene. Sufficient --and solid -- evidence, in particular for risk management purposes, is already available \textit{in the tail properties themselves}. An existential risk needs to be killed in the egg, when it is still cheap to do so.
Events of the last few months have shown that waiting for better data has generated substantial delays, causing thousands of deaths \textit{and} serious economic consequences. 

Secondly, unreliable data\footnote{Many of those complaining about the quality of data and asking for more evidence before taking action, even in extremely risky situations, rarely treat the inputs of their predictive models as imprecise \cite{CiTa,Viertl}, stressing them, and performing serious robustness checks of their claims.}--or any source of serious uncertainty--should, under some conditions, make us follow the "paranoid" route. More uncertainty in a system makes precautionary decisions more obvious. If you are uncertain about the skills of the pilot, you get off the plane \textit{when it is still possible to do so}. If there is an asteroid headed for earth, should we wait for it to arrive to see what the impact will be? We might counter that there were asteroids in the past that had devastating impacts, and besides we can calculate the physics. 
The logical fallacy runs deeper: "We did not see this particular asteroid yet" misses the very nature of the power of science to generalize (and classify), and the power of actions to possibly change the outcome of events. 
Similarly, if we had a hurricane headed for Florida, a statement like "We have not seen this hurricane yet, perhaps it will not be like the other hurricanes!" misses the essential role of risk management: to take preventive actions, not to complain ex post. And if people take action boarding up windows, and evacuating, the claim "look it was not so devastating", that someone might afterwards make, should be considered closer to a lunatic conspiracy fringe than scientific discourse.

By definition, evidence follows and never precedes rare impactful events. Waiting for the accident before putting the seat belt on, or evidence of fire before buying insurance would make the perpetrator exit the gene pool. Ancestral wisdom has numerous versions such as \textit{Cineri nunc medicina datur} (one does not give remedies to the dead), or the famous saying by Seneca \textit{Serum est cavendi tempus in mediis malis} (you don't wait for peril to run its course to start defending yourself).

However, just as there are frivolous lawsuits there are frivolous risk claims and, as we will see further down, we limit these precautionary considerations to a precise class of fat tailed multiplicative processes --when there is systemic risk. 

\begin{remark}[Fundamental Risk Asymmetry]
For matters of survival, particularly when systemic, and in the presence of multiplicative processes (like a pandemic), we require "evidence of no harm" rather than "evidence of harm." 
\end{remark}

\section*{\color{Brown} Technical Comments}
\subsubsection*{\color{Brown}The Law of Large Numbers (LLN) and Evidence}
In order to leave the domain of ancient divination (or modern anecdote) and enter proper empirical science, forecasting must abide by both evidentiary and probabilistic rigor. Any forecasting activity about the mean (or the parameter) of a phenomenon requires the working of the law of large numbers (LLN), guaranteeing the convergence of the sample mean at a known rate, when the number $n$ of observations increases. This is surely well-known and established, except that some are not aware that, even if the theory remains the same, the actual story changes under fat tails.

Even in front of the most well-behaved and non-erratic random phenomenon, if one claimed fitness or non-fitness of a forecasting ability on the basis of a single observation ($n=1$), he or she would be rightly accused of unscientific claim. Unfortunately, with fat-tailed variables that "$n=1$" error can be made with $n=10^6$. In the case of events like pandemics, even larger $n\to \infty$ can still be anecdotal. 

\begin{remark}[LLN and speed of convergence]
Fat-tailed random variables with tail exponent $\alpha \leq 1$ are simply not forecastable. They do not obey the LLN, as their theoretical mean is not defined, so there is nothing the sample mean can converge to. But we can still understand several useful tail properties. 

And even for random variables with $1<\alpha \leq 2$, the LLN can be extremely slow, requiring an often unavailable number of observations to produce somehow reliable forecasts.
\end{remark}

As a matter of fact, owing to preasymptotic properties, a conservative heuristic is to consider variables with $\alpha\leq 2.5$ as not forecastable in practice. Their sample mean will be too unstable and will require way too much data for forecasts to be reliable in a reasonable amount of time. Notice in fact that $10^{14}$ observations are needed for the sample mean of a Pareto "80/20", with $\alpha \approx 1.13$, to emulate the gains in reliability of the sample average of a 30-data-points sample from a Normal distribution \cite{TaStat}.

Assuming significance and reliability with a low $n$ is an insult to everything we have learned since Bernoulli, or perhaps even Cardano. 	

Also notice that discussing the optimality of any alarm system \cite{Amaral, Lindgren,Svensson} trying to perform predictions on averages would prove meaningless under extremely fat-tails, i.e. when $\alpha \leq 2$, that is when the LLN works very slowly or does not work. In fact, even when the expected value is well-defined (i.e. $1<\alpha<2$), the non-existence of the variance would affect all the relevant quantities for the verification of optimality \cite{deMare}, from the size of the alarm to the number of correct and false alarms, from the probability of detection of catastrophes to the chance of undetected events. For all these quantities, the naive sample estimates commonly used would prove misleading. A solution could be the implementation on EVT-based approaches, possibly with the additional tools of \cite{CiTaQF} or \cite{Neslehova}, but at this stage nothing similar exists, to the best of our knowledge.

For this and other reasons specified later, the application of a non-naive precautionary principle \cite{Prec} appears to be the viable solution in front of potentially existential risks.

\subsubsection*{\color{Brown} Science is about understanding properties, not forecasting single outcomes}

Figures \ref{lognormal} and \ref{pareto} show the extent of the problem of forecasting the average (and so other quantities) under fat tails. Most of the information is away from the center of the distribution. The most likely observations are far from the true mean of the phenomenon and very large samples are needed for reliable estimation. In the lognormal case of Figure \ref{lognormal}, 85\% of all observations fall below the mean; half the observations even fall below 13\% of the mean. In the Paretian situation of Figure \ref{pareto}, mimicking the distribution of pandemic deaths, the situation gets even worse: the mean is so far away that we will almost never observe it. It is therefore preferable to look at other quantities, like for example the tail exponent.

In some situations of fast-acting LLN, as (sometimes) in physics, properties can be revealed by single predictive experiments. But it is a fallacy to assume that a single predictive experiment can actually validate any theory; it is rather a single tail event that can falsify a theory.

Sometimes, as recently shown on the \textit{International Journal of Forecasting} by one of the authors \cite{taleb2020statistical}, a forecaster may find a single quantity that is actually forecastable, say the survival function. For $n$ observations a tail survival function has an error of $o(\frac{1}{n})$, even when tail moments are not tractable, which is why many predict binary outcomes--as with the "superforecasting" masquerade. In \cite{TaStat}, it is shown how--paradoxically--the more intractable the higher moments of the variable, the more tractable the survival function becomes. Metrics such as the Brier score are well adapted to binary survival functions, though not to the corresponding random variables. That is why survival functions are essentially useless for risk management purposes. In insurance, for instance, one never uses survival functions for hedging, but rather expected shortfalls--binary functions are reserved to (illegal) gambling.\footnote{The main problem is that the conditional expectation is not convergent: $\lim_{K \to \infty} \frac{1}{K} \mathbbm{E}(X|X>K) >1$, see \cite{taleb2020statistical} for a lengthy discussion.}

\subsubsection*{\color{Brown} We do not observe properties of empirical distributions}
A commentator (Andrew Gelman) \cite{Gelmanblog}  wrote "The sad truth, I'm afraid, is that Taleb is right: point forecasts are close to useless, and distributional forecasts are really hard." 

The problem is actually worse. In fact, distributional forecasts are more than hard--and often uninformative. Building so-called empirical distributions by survival functions does not reveal tail properties since it will necessarily be censored and miss tail observations --those that under very fat tails (say $\alpha \leq 2$) harbor not most, but literally all of the properties \cite{TaStat}. In other words, probabilities are thin-tailed (since they are bounded by $0$ and $1$) but the corresponding payoff is not, so small errors in probability translate into large changes in payoffs. However, as further discussed in \cite{CiTaNat}, the tail parameters are themselves thin-tailed distributed, hence reveal their properties rather rapidly. Simply, tail parameters extrapolate--while survival functions don't--and methods to measure the tail are quite potent.\footnote{This also relates to the superforecasters masquerade mentioned earlier: building survival functions for tail assessments via sports-like "tournaments" as in \cite{Tetlock}, instead of using more rigorous approaches like EVT, is simply wrong and violates elementary probability theory. }

\begin{figure}[h!]
\includegraphics[width=\linewidth]{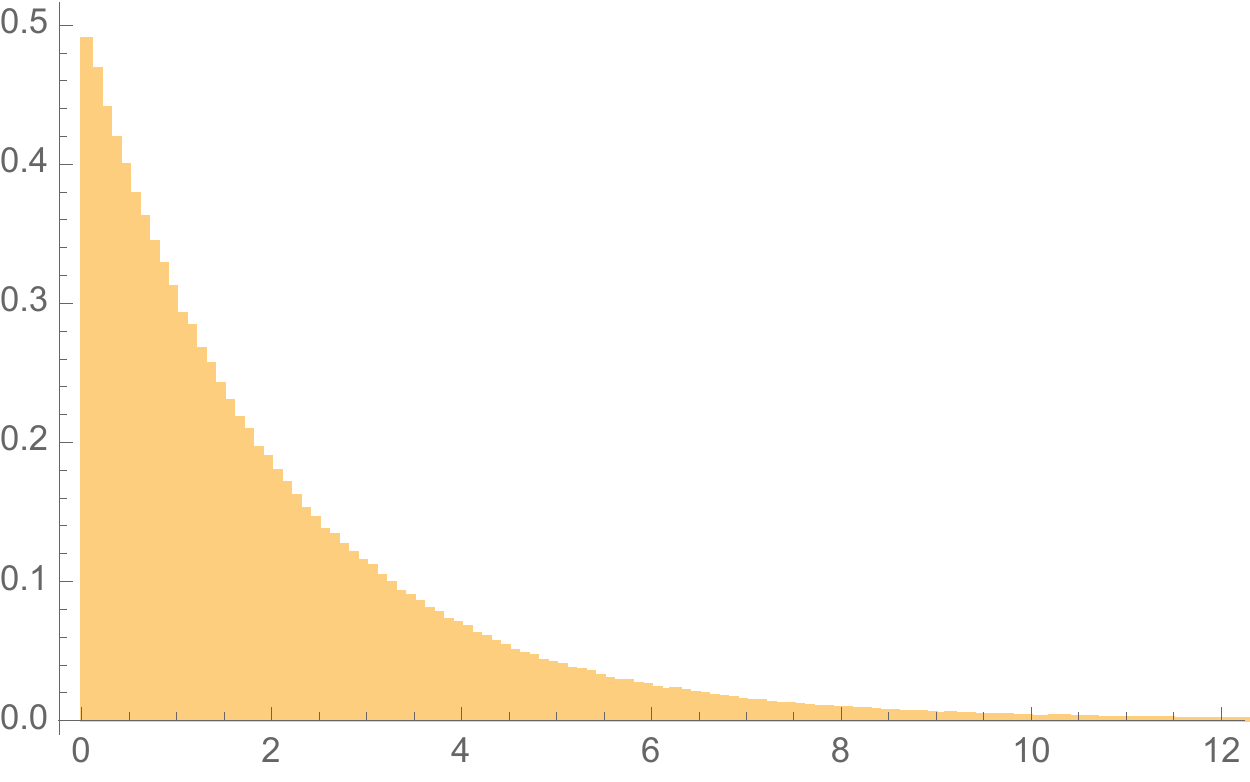}
\includegraphics[width=\linewidth]{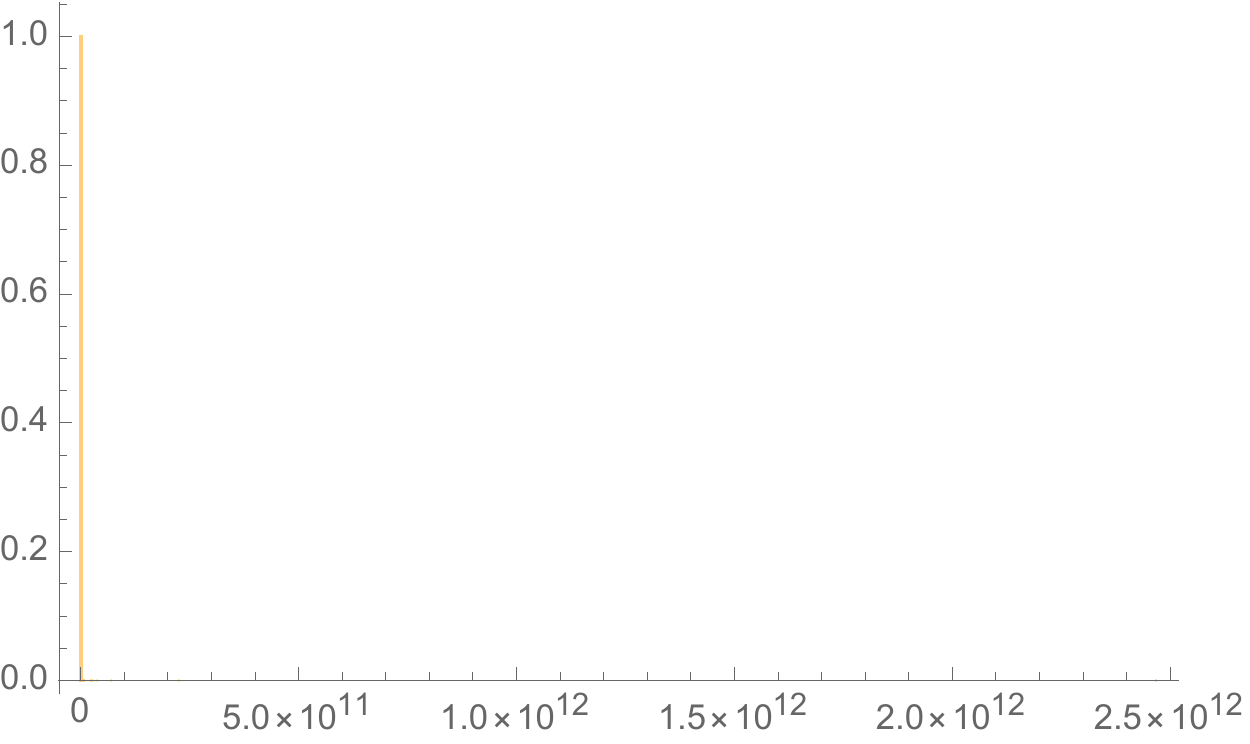}
 \caption{Above, a histogram of $10^6$ realizations of $r$, from an exponential distribution with sole parameter $\lambda=\frac{1}{2}$. Below, that of $X= e^r$. We can see the difference between the two distributions. The sample kurtosis are $9$ and $10^6$ respectively (in fact it is infinite for the second) --all values for the second one are dominated by a single large deviations.}\label{pareto}
\end{figure}

\subsubsection*{\color{Brown} Uncertainty goes one way; errors in growth rates induce biases and massive fat tails for the quantity of interest}
 
Consider the simple model 
$$X_t= X_0 \mathrm{e}^{r (t-t_0)},$$ 
where $X_t$ represents the quantity of interest (say the number of fatalities in pandemics) between periods $t_0$ and $t$,  $$r= \frac{1}{(t-t_0)} {\int_{t_0}^t r_s ds}$$ and $r_s$ is an instantaneous rate. 

Using the histograms of $r$ and $X$, Figure \ref{pareto} shows something fundamrntal: a well-behaved distribution, that of $r$, may lead to an untractable one, that of $X$; furthermore, the more volatile $r$, the more downward-biased your observation of the mean of $X$.

Implication: one cannot naively translate between the rate of growth $r$ and $X_T$, because errors in $r$ could be small (but surely not zero), but their impact will be explosive on $X$, because of exponentiation.  

Simply, if $r$ is exponentially distributed (or part of that family), $X$ will be power law. The tail $\alpha$ is a direct function of the variance: the higher the variance of $r$, the thicker the tail of $X$.

\begin{remark}[Errors in Exponential Growth]
1) Errors in growth rates of a disease increase the fatness of tails in the distribution of fatalities.\\
\noindent2) Errors in growth rates translate, on balance, into higher expected casualties.
\end{remark}

We note that in the context of dynamical systems an exponential dynamics is defined as chaotic \cite{Lorenz}.  While the study of chaos often considers systems with fixed parameters and variable initial conditions, the same sensitivities arise due to variations in parameters; in this case, the value of contagion rate ($R$) and the social behaviors that affect it. Indeed this means that by changing human behavior, the dynamics can be strongly affected, thus allowing for the opening of opportunities for extinction.\footnote{One of the authors has shown \cite{BYadded} that with increasing global transportation there is a phase transition to global extinction with probability 1. This indicates that historical distributions don't account for the severity or frequency of current or future extreme events because the fat tailed distributions themselves are coalescing to unit probability extreme events over shorter time intervals due to global changes in societal behaviors. During this process the probability distributions for events in any time interval becomes progressively more weighted to large scale events. Thus historical decadal or century intervals between pandemics are inadequate descriptors of current risk. }

\subsubsection*{\color{Brown}Never cross a river that is 4 feet deep on average } Risk management (or policy making) should focus on tail properties and not on the body of probability distributions. For instance, The Netherlands have a policy of building and calibrating their dams and dykes not on the average height of the sea level, but on the extremes, and not only on the historical ones, but also on those one can expect by modelling the tail using EVT, via semi-parametric approaches \cite{deHaan,Embrechts}.

\subsubsection*{\color{Brown} Science is not about safety} Science is a procedure to update knowledge; and it can be wrong provided it produces interesting discussions that lead to more discoveries. But real life is not an experiment.  If we used a p-value of $.01$ or other methods of statistical comfort for airplane safety, few pilots and  flight attendants would still be alive. For matters that have systemic effects and/or entail survival, the asymmetry is even more pronounced.

\subsubsection*{\color{Brown} Forecasts can result in adjustments that make forecasts less accurate}

It is obvious that if forecasts lead to adjustments, and responses that affect the studied phenomenon, then one can no longer judge these forecasts on their subsequent accuracy. Yet the point does not seem to be part of the standard discourse on COVID-19.

By various mechanisms, including what is known as Goodhart's law \cite{Strathern}, a forecast can become a target that is gamed by participants--see also the Lucas' critique applying the point more generally to dynamical systems. In that sense a forecast can be a warning of the style "if you do not act, these are the costs" \footnote{For instance Dr. Fauci's warning that the number of (verified) infections could reach 100K per day (\textit{New York Times}, June 30, 2020) should not be interpreted as a forecast to be judged according to its accuracy; rather a signal about what could happen should one avoid taking action.}.

More generally, any game theoretical framework has an interplay of information and expectation that causes forecasts to become self-canceling. The entire apparatus of efficient markets--and modern economics--is based on such self-canceling aspect of prediction, under both rational expectations and an arbitrage-free world.

\section*{\color{Brown}Remarks Specific to Ioannidis et al.}\label{answer}

\subsubsection*{\color{Brown}Systemic risks vs individual risks}

A fundamental problem, in both \cite{Ioannidisfirstpaper} and \cite{Ioannidisblog}, lies in ignoring scaling: systemic risks do not resemble (even qualitatively) individual risks. The macro and the micro-properties of contagious events, given their infective multiplicative nature, don't map directly onto one another.

Ioannidis et al. \cite{Ioannidisfirstpaper} write: "the average daily risk of dying from coronavirus for a person <65 years old is equivalent to the risk of dying driving a distance of 13 to 101 miles by car per day during that COVID-19 fatality season in 17 of the 24 hotbeds (...) For many hotbeds, the risk of death is in the same level roughly as dying from a car accident during daily commute."

 Even if Ioannidis et al.'s computation were to  hold true for one individual (it does not), conditionally on an excess of $10^3$ of such individuals dying, the probability that the cause of death is COVID-19 and not a car accident converges to 1. When you die of a contagious disease, people around you are at risk of contagion, and they can then infect other people, in a cascading effect. 
 It is quite elementary: car accidents are not contagious, while COVID-19 is. You cannot conflate the two objects: one is additive in the aggregate, the other is multiplicative. In \cite{TaStat}, it has been shown that this is a severe error, leading to macroscopic blunders\footnote{Note that this is also a typical example of "size fallacy," in which different risky events are compared just on the basis of their probabilities of occurrence, without caring about their different nature \cite{Hansson}.}.

\begin{remark}[Additive vs. Multiplicative Risks (Scaling of Probabilities)]

	Under multiplicative effects the risks for a collective do not scale up from the risks of an individual. Trivially, systemic risks can be extreme, where the individual ones are low, or vice-versa.
\end{remark}

\subsubsection*{\color{Brown}Trade-offs and Ergodicity} One could say: panic saves lives, but at what economic price? Let us put aside ethical arguments, and answer it, ignoring for a moment the value of human life. 

The fact is that some classes of (systemic) risks require being killed in the egg, also from an economic point of view. The good news is that there are not so many--but pandemics as we said fall squarely within the category.

The "dismal" theorem \cite{Weiz} mentioned earlier tells us that it is an error to use trade-off analysis under existential risk. There have been many proofs of similar arguments on grounds of ergodicity, well-known by insurance companies since Cram\`er: simply, you cannot use naive B-school costs-benefit analyses for Russian roulette, because of the presence of an absorption barrier \cite{Incerto}. But one should not blame Ioannidis et al. \cite{Ioannidisfirstpaper} for this error in reasoning: it has been shown to be unfortunately prevalent in the decision-science literature \cite{Peters}.

\begin{remark}[Ruin Problems]
Traditional cost-benefit analysis fails to apply to situations where statistical averages are unreliable, if not invalid.
\end{remark}

Moreover, it is not correct to assume, more or less implicitly, that a disease brings no or little costs, while mitigation is burdensome. There are indeed severe nonlinearities at play.  

First of all, risk is beyond the simple and direct disease-specific mortality rate. In fact, letting the disease run above a certain threshold would compound its effect (in an explosive manner), because of the saturation of services, causing for example the displacement of other patients, many in potentially critical conditions; something that we have seen happening in the Region of Lombardy in Italy, in New York City, and elsewhere for several weeks during the spring of 2020 \cite{Soreide}. Furthermore, for survivors the illness itself represents a large economic drain, be it only from lost working hours, not counting the costs of hospitalization. And for every severe infection, there is an unspecified number of morbidities, with unknown (but definitely larger than zero) additional mortality and long term costs for the health system \cite{Ackerman,Soreide}, as it has been the case for other diseases like SARS \cite{Ngai}.\footnote{\textbf{Geronticide:} This discussion does not even cover the ethical discussion of trade-offs and their inapplicability in some domains, perhaps the most central discussion. At what price will you kill your parents/grandparents? A million dollars? Ten million? A billion? Furthermore, the fact that older people are more vulnerable to the disease brings considerations of geronticide (senicide): one misses that the silver rule \cite{Incerto} commands treating older generations under a moral liability, as one wishes to be treated by the next generation. Letting the disease run through older generations violates the interdicts on geronticide and intergenerational obligations. The fact that your parents did not sacrifice their own parents creates an obligation to not sacrifice them; your children will spare you in turn, under the same rule.}

 \begin{remark}[False Dichotomies]
One should not treat the economy and the disease as separate independent items, particularly by viewing a naive trade-off between economic costs and pandemic mitigation. 
\end{remark}

Moreover, never underestimate consumers' (nonlinear) behavior. When risks are visible (and a pandemic definitely is), people tend to modify their behavior, rationally or not, also switching to alternatives, with nonlinear effects on the businesses concerned \cite{Rose}. This is the reason why the airline industry in the U.S. manages to have fewer than $1$ fatal crash in $25 \times 10 ^{6}$ flights (and aims at an even more favorable ratio). One may claim that it is irrational to spend so much of our resources mitigating plane crashes, but airline companies know that, in case of fewer checks and efforts, consumers would then probably switch to other companies, if not directly to other types of transportation.

Take the hospitality industry. Unless there is once again comfort on the part of the public, restaurants and hotels will be unprofitable. The rule of thumb in NYC is that a drop of 15\% in revenues is sufficient to make a restaurant shutter permanently; there has been a large drop in restaurant attendance in Sweden where the state did not enforce lockdowns, owing to a high rate of voluntary self-isolation \cite{Kamerlin}.

The United States (and many other countries worldwide) have spent trillions of dollars on sophisticated weaponry in the past decades, to counter \textit{uncertain} threats. It would be a good idea to question these expenditures first, before doubting the spending to stave off certain pandemics. 

Likewise, it would be a good idea to question first the excessive burden on Western economies, particularly the U.S., of measures taken to ensure workplace and transportation safety which, we saw, are driven by the legal system and the tort mechanisms.

\begin{remark}[Domain Dependence]
	It is not rational to worry about pandemic costs (extremely fat-tailed exposure), while not \textit{also} questioning other sizable insurance-style expenditures for transportation and workers safety.
\end{remark}

It is therefore incorrect to claim that it is the authorities' response to the pandemic that caused unemployment in the transportation or hospitality industries. As a matter of fact, the arguments proposed by two of the authors \cite{Prec}, last January 2020, were aimed at lowering the economic effect of the pandemic: prevention is orders of magnitude cheaper than the cure--recall that \textit{sed prior est sanitas quam sit curatio morbi.} 

We note that many comments of the type "the pandemic has caused \textit{only} 640K fatalities" (as of July 25, 2020) simply ignore the fact that, in practically every location subject to the pandemic, there has been local or governmental action to mitigate it--we do not consider the counterfactual of "what if" because it is not visible.

\begin{remark}[Economic Fragility]
The argument in \cite{Incerto} is that we live in an over-optimized environment, in which a slight drop in sales or a change in consumer preferences may cause wild interlocking industry collapses. This nonlinearity is similar to "large a movie theater with a very small door at the times of fire." 

It is more cogent to blame the over-optimized economic structure than the general reaction to the disease.
\end{remark}

\begin{figure*}[h!]
\includegraphics[width=\textwidth]{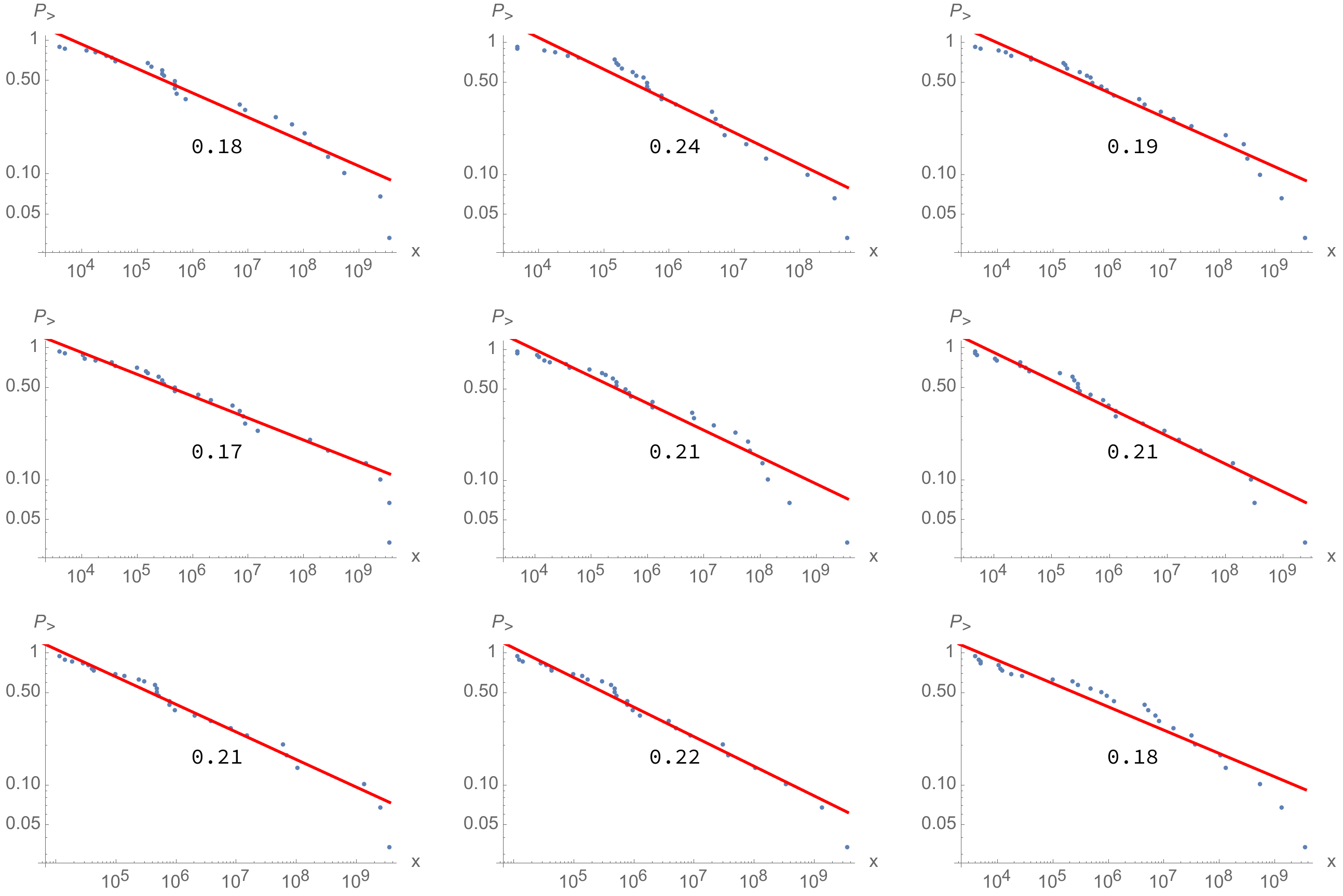}
\caption{Zipf plots (log-log plots of the empirical survival function $P_{>}$) for nine random selections of 30 out of the 72 pandemics in \cite{CiTaNat}. The number in the center represents the a naive (OLS) estimate of the tail parameter $\alpha$, readable as the absolute slope of the red negative line. The values of $\alpha$ appear to be stable notwithstanding the sampling, signalling the robustness of the approach and the inconsistency of the "selection bias" critique. The values are also in line with the more rigorous EVT-based findings of \cite{CiTaNat}.} \label{zipf}
\end{figure*}

\subsubsection*{\color{Brown} Early Mitigation and Economic History} 
We note here that while the Great Plague took place in the fourteenth century, quarantines were enforced five centuries later as economies understood they could not afford recurrences. Between the Habsburg and the Ottoman Empire, there were \textit{lazarettos} along the border, and every active Mediterranean port enforced quarantines for travelers along the expanded silk road, while pilgrim routes were subjected to similar measures. For instance, in the 1830s, in \textit{the Count of Monte Christo}, a traveler from Paris to Ioanina (where Prof. Ioannidis was previously located), had to spend four days in quarantine to get there, while there was no particular threat of disease. In fact, the novelist was underestimating, for historian records show mandatory nine days for ordinary travelers and fifteen days for merchants according to \cite{Sarilyildiz, Roberts}.  Economies adapted to early mitigation throughout the centuries preceding our era. Furthermore, the Ottoman Empire has ready \textit{lazaretos} for additional quarantining along specified stations at the first signs of a pandemic.

Mitigation has another effect: to delay and temporize, while we can understand the properties of the disease. While initial treatments are under high opacity, later treatments allow for gains of collective experience\footnote {Aside from considerations of geronticide, when the costs of the Swedish experiment are finally told, one of the factors will be the early loss of life when later (current) medical practice would have saved them even before a vaccine.} .


\subsubsection*{\color{Brown} Selection Bias and Class of Events} 
In \cite{Ioannidisblog}, the authors erroneously maintain that choosing tail events as done by \cite{CiTaNat} is "selection bias". Actually, the standard technique there used is the exact opposite of selection bias: in EVT, one purposely focuses on extremes to derive properties that influence the outcomes, especially from a risk management point of view. One could more reasonably argue that the data in \cite{CiTaNat} do not contain all the extremes, but, by jackknifing and bootstrapping the data, the authors actually show the robustness of their results to variations and holes in historical observations: the tail index $\alpha$ is consistently lower than 1. In Figure \ref{zipf} a simple illustration is given, showing that one can be quite radical in dealing with the uncertainty in pandemic fatalities, and still find out that the findings of \cite{CiTaNat} hold true.

When the authors in \cite{Ioannidisblog} state that "Tens of millions of outbreaks with a couple deaths must have happened throughout time," to support their selection bias claim against \cite{CiTaNat}, they seem to overlook the fact that the analysis deals with pandemics and not with a single sternutation. The class of events under considerations in \cite{CiTaNat} is precisely defined as "pandemics with fatalities in excess of $1K$," and their dataset likely contains most (if not all) of them. Worrying about many missing observations in the left tail of the distribution of pandemic deaths is thus misplaced.

\subsubsection*{\color{Brown} Conditional information} One may be entitled to ask: as we get to know the disease, do the tails get thinner? Early in the game one must rely on conditional information, but as our knowledge of the disease progresses, shouldn't we be allowed to ignore tails?

Alas, no. The scale of the pandemic might change, but the tail properties will remain invariant. Furthermore, there is an additional paradox. If one does not take the pandemic seriously, it will likely run wild (particularly under the connectivity of the modern world, several orders of magnitude higher than in the past \cite{AlbertBarabasi}). And diseases mutate, increasing or decreasing in both lethality and contagiousness. The argument would therefore resemble the following: "we have not observed many plane crashes lately, let's relax our safety measures".

Finally, we conclude this section with an encouraging point: fat tails do not make the world more complicated and do not cause frivolous worries, to the contrary. Understanding them actually reduces costs of reaction because they tell us what to target--and when to do so. Because network models tend to follow certain patterns to generate large tail events \cite{AlbertBarabasi, Garibaldi}, in front of contagious diseases wisdom in action is to kill the exponential growth in the egg via three central measures 1) reducing super-spreader events; 2) monitoring and reducing mobility for those coming from far-away places (via quarantines); 3) looking for cheap measures with large payoffs in terms of the reduction of the multiplicative effects (e.g. face masks\footnote{Most of the trillions spent could have been saved if authorities understood the double nonlinearities in face masks: 1) the compounding effect of both parties having protection, 2) the nonlinearity of the dose response with disproportional drop in the probability of infection from a reduction in viral load \cite{TalebMasks}.}). Anything that "demultiplies the multiplicative" helps \cite{TalebMasks}.

 Drastic shotgun measures such as lockdowns are the price of avoiding early traveler quarantines and border monitoring; they can  be --\textit{temporarily and cum grano salis} --of help, especially in the very early stages of the new contagious disease, when uncertainty is maximal, to help isolating and tracing the infections, and also buying some time for understanding the disease and the way it spreads. Indeed such drastic and painful measures can carry long-lasting damages to the system, not counting an excessive price in terms of personal freedoms. 
 
 But they are the price of not having a good coordinated tail risk management in place --to repeat, border monitoring and control of superspreader events being the very first such measures. And lockdowns are the costs of ignoring arguments such as increased connectivity in our environment and conflating additive and multiplicative risks.
 

\section*{***}
To conclude, as the trader lore transmitted by generations of operators goes, "if you must panic, it pays to panic early." 

The Ottoman Empire integrated Byzantine knowledge accumulated since at least the Plague of Justinian; it is sad to see ancient cultures more risk-conscious, better learners from history, and economically more effective than modern governments. They avoided modern "evidence based" reductions that, as we saw, are insulting to both science and wisdom. And, had it not been for such a collective ancestral risk-awareness and understanding of asymmetry, we doubt that many of us would be here today.

Now, what did we learn from the pandemic? That an intelligent application of the precautionary principle \cite{Prec} consists in formulating decisions that are wise in both foresight and hindsight. Here again, this is ancient: it maps to Aristotle's \textit{phronesis} as presented in his \textit{Nichomachean Ethics}.

\end{document}